# AI for Quality Assurance in the Operating Room

**Pietro Mascagni[1, 2], Lalith Sharan[2, 3], Deepak Alapatt[4], Nicolas Padoy[2, 3]**

[1]Fondazione Policlinico Universitario Agostino Gemelli IRCCS, Gemelli, Rome, Italy
[2]Institute of Image-Guided Surgery, HU-Strasbourg, Strasbourg, France
[3]University of Strasbourg, CNRS, INSERM, ICube, UMR7357, Strasbourg, France
[4]Scialytics SAS, Strasbourg, France

**Corresponding author:**
Pietro Mascagni, MD, PhD
Università Cattolica del Sacro Cuore
Largo Francesco Vito, 1, 00168 Roma RM, Italy
Telephone: +39 3922736975
Email: pietro.mascagni@ihu-strasbourg.eu

Surgical outcomes depend not only on patient factors and postoperative care but are also strongly influenced by the quality of the operation itself. Yet, for much of modern surgery, intraoperative quality has been assessed indirectly through outcomes and operative reports. The increase in minimally invasive procedures inherently guided by endoscopic video, together with advances in artificial intelligence, creates an unprecedented opportunity to systematically observe, measure, and improve surgical care. This chapter introduces AI-enabled Surgical Quality Assurance as a framework for using surgical data to support continuous assessment and improvement in the operating room. We first review existing approaches to surgical safety, from system-level interventions to procedure-specific standards. We then describe how AI can transform intraoperative video into clinically meaningful information, including recognition of anatomy, instruments, workflow, surgical actions, quality criteria, adverse events, and critical moments. Finally, we outline the major challenges that must be addressed before these systems can deliver routine clinical value, including representative data collection, robust validation, workflow integration, regulation, liability, privacy, and equitable access. Rather than replacing surgical judgment, AI for quality assurance should be understood as a set of tools for augmenting the surgical team, scaling expert review, and helping surgery evolve toward a learning system in which intraoperative care is continuously observed, assessed, and improved.

## 1 Introduction

More than 300 million surgical procedures are performed worldwide every year (Weiser et al. 2015), and one in seven of those patients experiences a postoperative complication (International Surgical Outcomes Study Group 2016). The human cost of surgical adverse events is hard to overstate: complications increase mortality (Fowler et al-. 2022), erode quality of life, and weigh on the surgeons and teams involved, often referred as "second victims". The financial cost is correspondingly large: complications increase per-case hospital expenditure by roughly 200% in mature health systems (Healy et al. 2016; Watson et al. 2025). These figures place the intraoperative phase of surgical care – the focus of this chapter, among the most consequential and least understood components of modern medicine.

Modern surgery is a highly effective sociotechnical process whose outcomes depend on the interplay between technical skill, clinical judgement, team dynamics, equipment, and patient factors (Mascagni and Padoy 2021). Historically, the field has had few direct ways to study what actually happens between skin incision and closure. Operative reports are dictated retrospectively by the operating surgeon and capture a sanitized summary of events. Postoperative outcomes are multifactorial and reveal problems too late to act on them. Quality and safety have therefore long been inferred indirectly, by aggregating outcomes at the level of the surgeon, the hospital, or the registry, and by deducing which intraoperative behaviors might explain the variation observed (Birkmeyer et al. 2013). This indirect lens is informative but coarse. It tells us that surgeons differ, but not how, and it offers little guidance on which specific intraoperative behaviors to reinforce or correct.

The advent of minimally invasive surgery (MIS) has changed this landscape. Laparoscopic and robotic procedures are guided by an endoscopic camera that captures a continuous, high-resolution stream of the operative field. This stream is a native output of the procedure, and its collection does not require any additional device or sensor. As MIS has spread across abdominal, thoracic, urological, gynecological, skull base, and cardiothoracic surgery, the operating room (OR) has become one of the richest sources of digital data in medicine. For the first time, intraoperative care can be observed directly rather than reconstructed from outcomes, and this shift is propelling a new wave of interest in surgical safety from clinicians, researchers, educators, payers, and regulators alike.

Consequently, surgical communities have begun to define and audit explicit standards of intraoperative quality, for example, the Critical View of Safety (CVS) for laparoscopic cholecystectomy (Strasberg et al 1995). Similar criteria are being formalized for inguinal hernia repair, appendectomy, colorectal resection, and other procedures (Daes and Felix 2017, Subramanian et al. 2012, Degiuli et al. 2023). Video-based assessment (VBA) has additionally matured into a credible method for measuring intraoperative performance, with structured competency assessment tools (CATs) developed and validated for use in randomized trials (Tsai et al. 2019) and surgical training (Miskovic et al. 2013). Furthermore, multi-modal recording platforms have brought the wider OR into view, enabling systematic study of distractions, communication failures, and team behaviors that affect outcomes (Goldenberg et al. 2017).

However, multiple practical considerations exist in realizing these standards. Surgeons frequently fail to apply the standards in practice. Video-based audits of laparoscopic cholecystectomies found that the CVS was correctly achieved in 9% to 16% of procedures despite its universal recommendation (Korndorffer et al. 2020, Mascagni et al. 2021). Secondly, most validated standards and CATs cover a handful of procedures; the methodology for defining new ones, based on cognitive task analysis, expert interviews, and Delphi consensus, is rigorous but slow, and producing a new tool from scratch can take years (Tsai et al. 2019; Miskovic et al. 2013). Replicating this effort across the long tail of surgical procedures requires systematic ways to look at data and codify expert knowledge. Finally, even where standards and CATs exist, applying them at the scale of routine surgical practice currently requires expert assessors reviewing video manually, which is neither sustainable nor objective enough to support quality improvement over time. Closing these implementation, replicability, and scale gaps likely requires automated tools for systematically collecting surgical videos, analyzing them to extract insights at scale, and providing real-time intraoperative assistance.

Surgical Data Science (SDS), an emerging discipline at the intersection of surgery and computer science, addresses these bottlenecks (Maier-Hein et al. 2017, 2022). Its premise is that the digital data generated during surgery, in particular videos, can be modelled with artificial intelligence (AI) to extract clinically meaningful information at scale. AI offers a path to real-time assistance that might help closing the implementation gap, to automated analysis that helps codify and replicate intraoperative standards, and to continuous monitoring that brings closed-loop quality assessment and improvement within reach. We hence use the term Surgical Quality Assurance (SQA) to describe the systematic use of surgical videos and analytics to sense, analyze, and assist surgeons in the operative room, with the goal of measuring and improving the quality and safety of intraoperative care.

This chapter examines how AI can support SQA to turn the OR from a highly consequential but scarcely optimized component of modern healthcare to a learning system that continuously improves for the benefit of all stakeholders, with the patient at the center. We begin by reviewing existing models of surgical safety in the OR, distinguishing between checklist-style interventions derived from outcomes-based reasoning and newer generation of practices grounded in direct observation of intraoperative video. In parallel, we outline the role played by non-technical skills and OR team dynamics. We then articulate the SQA vision and its operational requirements. The middle of the chapter introduces the technical fundamentals of AI for endoscopic video analysis and surveys current applications across surgical skill, procedural quality, adverse event detection, and event anticipation. We close with an honest assessment of the road ahead, organized around the data, methods, regulatory, and governance challenges that stand between today's proof-of-concept work and routine clinical value.

## 2 Operating Room Safety

Practitioners as well as innovators have always been keen to improve surgical safety. However, what has changed over the past three decades is the source of evidence to inform it. Early interventions were derived from indirect observations and statistical reasoning on outcomes; more recent ones are grounded in direct observations of intraoperative events. The two approaches are complementary, and the trajectory from the first to the second is itself part of the story of how the field is moving from inference to measurement.

### *2.1 Outcomes-based inference: the WHO checklist and system-level safety*

The first wave of modern surgical safety interventions emerged from epidemiological, and human-factors analyses of postoperative outcomes. Adverse event registries, malpractice claim reviews, and aviation-inspired error analyses converged on a striking observation: the underlying causes of harm in the OR were often communication failures, lapses in situational awareness, and breakdowns in coordination rather than purely technical errors (Helmreich 2000). The clearest response to this insight was the WHO Surgical Safety Checklist, introduced in 2008 as a 19-item tool structured around three time-bound checkpoints: before induction of anesthesia, before skin incision, and before the patient leaves the OR. In an eight-country pilot, its adoption was associated with a near-halving of postoperative mortality and a one-third reduction in major complications (Haynes et al. 2009). Subsequent meta-analyses confirmed an overall benefit, while highlighting substantial heterogeneity tied to the quality of implementation (Treadwell et al. 2014; Conley et al. 2011).

The checklist is informative both as an intervention and as a methodological example. It was developed by inferring causal mechanisms from outcomes data and codifying them as a set of static, peri-procedural prompts. It works at the level of the system, and its effects are generalizable. However, it does not work at the level of the single patient or procedure, and it is highly sensitive to local culture, such as the willingness of teams to engage substantively rather than performatively (Conley et al. 2011). The WHO Surgical Safety Checklist therefore illustrates both the value and the limits of classical, static safety interventions derived from outcomes-based inference: it captures broad categories of risk effectively but lacks the granularity necessary to inform the procedure itself.

## *2.2 Direct observation: the CVS and surgical process measures*

Minimally invasive surgery created the conditions for a different kind of safety work - one grounded in direct observation of the intraoperative scene. The earliest and most influential example concerns bile duct injuries (BDI) during laparoscopic cholecystectomy. Building upon the experience derived from open cholecystectomy and upon direct observations of the back then newly introduced laparoscopic approach, in 1995, Strasberg and colleagues described the Critical View of Safety (CVS) to prevent BDI (Strasberg et al. 1995). Eight years later, Way and colleagues conducted a landmark root-cause analysis of 252 BDIs, drawing on operative records and 22 videotapes of original procedures. The primary cause of error in 97% of cases was a visual perceptual illusion in which the common bile duct was mistaken for the cystic duct, with technical skill failures accounting for only 3% (Way et al. 2003). This analysis established that BDI is fundamentally a perceptual and cognitive event arising from intraoperative anatomical recognition, confirming the utility of a process to conclusively identify the hepatocystic anatomy, i.e., the assessment of CVS. Thanks to these observations and evidence derived from numerous case-series, the CVS is now formally recommended by multi-society clinical practices guidelines (Brunt et al. 2020).

Analogous video-derived standards have emerged across other procedures. In transabdominal preperitoneal (TAPP) inguinal hernia repair, the European Hernia Society and several national societies have codified the so-called critical view of the myopectineal orifice exposure, defining the anatomical landmarks that must be visualized before mesh placement to avoid injury to the iliac vessels and nerves (Daes and Felix 2017). In appendectomy, similar criteria have been proposed for safe identification of the appendiceal base and management of the mesoappendix (Subramanian et al. 2012). In gastric and colorectal resections, oncological quality measures such as adequate lymphadenectomy and intact specimen integrity are increasingly assessed by video or photo documentation (Degiuli et al. 2023).

## *2.3 Competency assessment: scaling video-based evaluation to complex procedures*

Where simple visual criteria such as the CVS suffice for relatively standardized procedures, more complex operations require richer assessment instruments. The push to develop them came from two communities. Surgical trialists running multicenter randomized trials needed to control for inter-surgeon variation, a confounder that can dominate the comparison of surgical interventions (Birkmeyer et al. 2013; McCulloch et al. 2009). Surgical educators needed granular, structured feedback to support competency-based training. Both have converged on video-based assessment using competency assessment tools (CATs) that can be applied reproducibly by multiple raters and correlated with outcomes.

The methodology for building such tools has stabilized over the past decade and typically follows four steps. The first is cognitive task analysis (CTA), often combined with hierarchical task analysis, in which expert surgeons are interviewed and observed to elicit the procedural decomposition, decision points, and error-prone steps that they use tacitly but rarely articulate (Sullivan et al. 2008). The second is broader expert consensus through a Delphi process: iterative rounds of anonymous surveys converge on which steps are mandatory, optional, or prohibited, and on the criteria distinguishing competent from poor performance (Diamond et al. 2014). The third is validation, including assessments of intra- and inter-rater reliability and correlation with outcomes. The fourth, increasingly explicit in recent years, is the consideration of how the resulting tool can be automated by AI to scale beyond manual review.

Two examples illustrate the model. For laparoscopic Nissen fundoplication, the Society of American Gastrointestinal and Endoscopic Surgeons (SAGES) developed a procedure-specific rating tool that decomposes the operation into key steps, each rated against explicit criteria, providing both formative feedback for trainees and a substrate for outcomes research. For total mesorectal excision, the international COLOR III randomized trial, comparing transanal TME with laparoscopic TME, established one of the most complete surgical quality assurance systems described to date. Tsai and colleagues used hierarchical task analysis to identify 101 task variations across 16 expert surgeons, applied a four-round Delphi process to reach consensus on 83 tasks, and constructed a CAT as a matrix of nine critical procedural steps and four performance domains, namely: Exposure, Execution, End-product Quality, and Adverse Events (Tsai et al. 2019). The resulting tool achieved a generalizability coefficient of 0.883 and was used not only for trial monitoring but as a criterion for trial entry: surgeons had to submit unedited videos and reach a minimum CAT score to participate. The COLOR III system shows what video-based quality assurance can achieve when implemented rigorously, and the same methodology is now being applied to procedures such as minimally invasive distal pancreatectomy and robotic gastrectomy where no validated framework yet exists.

## *2.4 The operating room as a whole: team dynamics and non-technical skills*

The work described above focuses on what happens at the surgical field level, captured by the endoscopic camera. A parallel and increasingly active line of research focuses on the OR as a whole, where team behavior, communication, and cognitive workload influence outcomes in ways that endoscopic video cannot capture. The conceptual foundations of this work were laid by the Non-Technical Skills for Surgeons (NOTSS) taxonomy, developed by Yule and colleagues, which decomposes intraoperative non-technical performance into four domains: situation awareness, decision-making, communication and teamwork, and leadership (Yule et al. 2008). NOTSS is now used internationally for assessment, training, and feedback, and has been shown to predict adverse events independently of technical skill (Jung et al. 2020).

Direct sensing of OR behavior is a more recent development. Grantcharov's group introduced the OR Black Box (Surgical Safety Technologies, Toronto, Canada), a multi-port recorder that synchronously captures audiovisual, physiological, and environmental data from the OR (Goldenberg et al. 2017). Manual analysis of OR Black Box recordings has shown a median of 20 errors per case, identified the most common distractions, and characterized resilience supports such as proactive task management and coaching (Goldenberg et al. 2017). Similarly, Beldi, Keller, and colleagues at the University of Bern developed and evaluated the StOP? protocol, a structured intraoperative briefing in which the surgeon pauses the team to summarize progress, share next steps, and surface concerns; analyses of OR audio and video have linked StOP? adoption with improved team communication and case awareness (Tschan et al. 2022; Chen et al. 2025a). In Boston, Zenati, Dias, and colleagues have studied cognitive workload during cardiac surgery using wearable sensors and machine learning, showing that heart rate variability and other physiological signals track meaningfully with case phase and workload, and proposing cognitive engineering frameworks to improve patient safety in cardiothoracic surgery (Zenati et al. 2020; Kennedy-Metz et al. 2020; Dias et al. 2022). Distractions and noise in the OR have similarly been linked to case difficulty and to patient outcomes (Peisl et al. 2024).

Two points are worth retaining from this body of work. First, team dynamics influence outcomes through mechanisms that are largely independent of those captured in the surgical field, and they require their own sensing modalities (audio, ceiling cameras, wearables) and their own intervention strategies (briefings, checklists, training, cognitive aids). Second, the toolkit for studying team dynamics is also moving from manual analysis toward automated AI-based methods. Recent work has begun to detect structured team communication events such as StOP? briefings directly from OR audio and video (Chen et al. 2025a), to estimate OR staff body pose and trajectories from low-resolution ceiling cameras (Chen et al. 2025b, Hamoud et al. 2024) and to model surgeon and team cognitive states from physiological signals (Kennedy-Metz et al. 2020). The endoscopic field and the OR are evolving to offer a holistic environment designed for sensing, analyzing, and assisting the surgical team.

## *2.5 A Vision for AI-Powered Surgical Quality Assurance*

The interventions described above have meaningfully improved surgical safety, but they share common limits. Each is implemented as a discrete, mostly manual practice, typically within well-defined research protocols or quality improvement initiatives. Notably, these interventions often lack a feedback system linking interventions with changes in practices and outcomes.

AI-Powered Surgical Quality Assurance reframes quality assessment and improvement as elements of a single continuous loop in which intraoperative care is systematically sensed, automatically analyzed against explicit surgeons-defined standards, and assisted in real time, with the same data building surgical intelligence feeding longer-term quality improvement (Fig. 1).

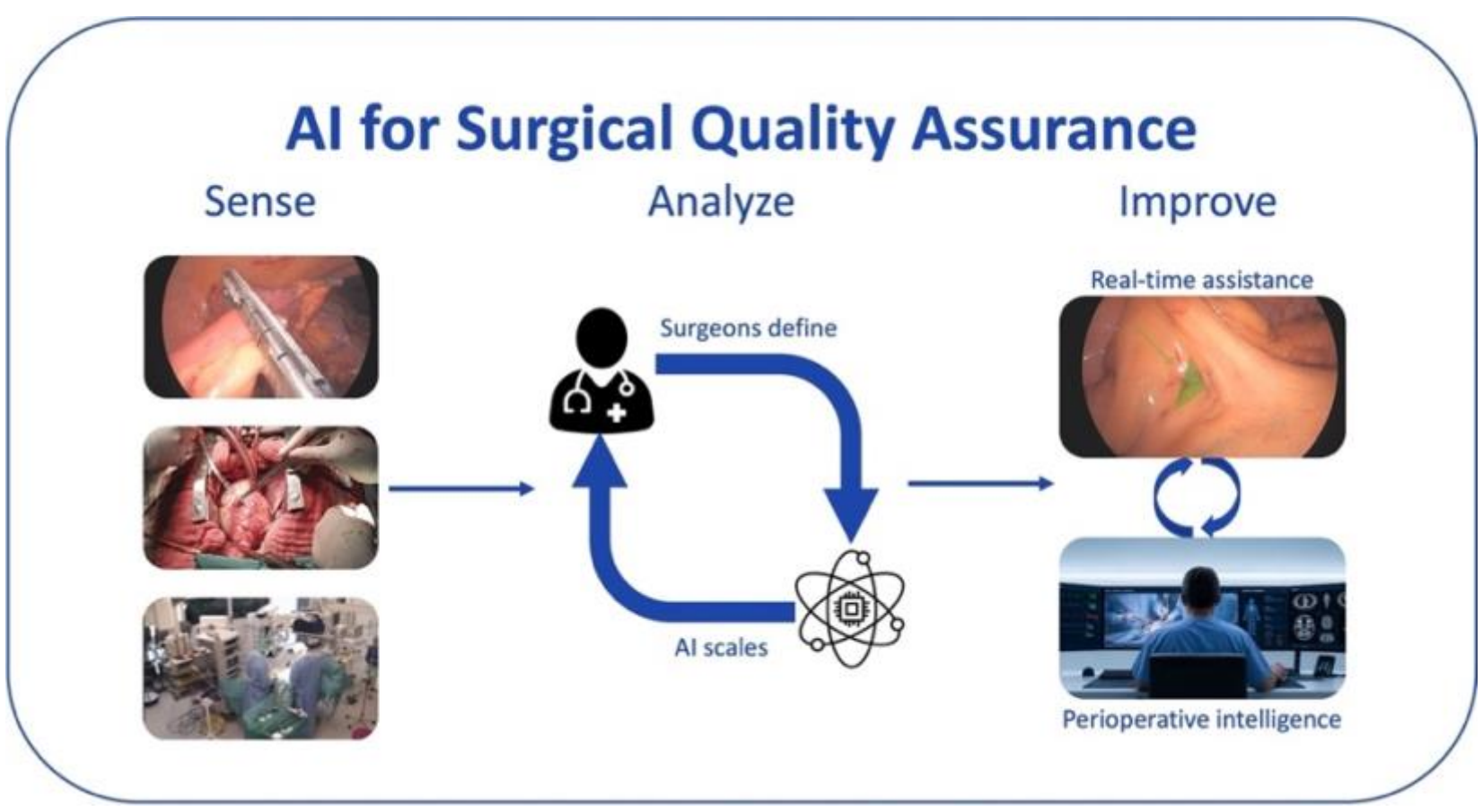

**Fig. 1 AI for Surgical Quality Assurance across three stages.** Sense**:** holistic capture of OR data. Analyze**:** surgeons define what matters while AI scales the analysis, combining domain expertise with large-scale processing. Improve**:** insights feed back into the OR as real-time assistance and perioperative intelligence.

Several operational requirements feed into this vision. Firstly, there is a need for a synchronized, standardized, and interoperable way to capture and record OR audio, video, robotic kinematics, device logs, anesthesia records, and physiological signals from the patient. Holistic data from the OR can help go beyond endoscopic video streams and can greatly enrich the information available for modelling. The infrastructure and standards for this kind of holistic sensing are emerging from both the MedTech and academic spaces (Mascagni and Padoy 2021). Secondly, we need to enable automated inference of relevant insights. Clinical events of interest, from anatomical structures and surgical phases to quality criteria and adverse events, must be extracted from these data reproducibly, at scale, and with sufficient timeliness to be acted on. Finally, the insights should feed into closed-loop assistance: the analyses must be delivered back to the surgical team in formats that support, rather than disrupt, intraoperative decision-making, and back to the institution in formats that facilitate training and continuous improvement.

As a result, surgical quality assessment can leverage these capabilities to measure what happens in the OR, for instance by assessing implementation rates of intraoperative key performance indicators such as the CVS or operational competency through CATs. Surgical quality improvement closes the loop by translating these measurements into targeted interventions: real-time prompts when a safety criterion is at risk of being missed, structured feedback to surgeons after their cases, identification of training needs, and refinement of standard operating procedures. The combination is what differentiates SQA from existing safety practices: it transforms surgical safety from a series of episodic, manually enforced checkpoints into a continuous, evidence-based discipline analogous to the quality systems that have matured in most consumer facing quality management systems.

AI is the technology that makes this vision possible. Without it, comprehensive sensing produces a flood of data that no clinician can manually digest; automated analysis remains a research aspiration and closed-loop assistance remains beyond reach. The remainder of this chapter describes the AI methods that underpin SQA and surveys what has already been demonstrated.

## 3 Artificial Intelligence for Endoscopic Video Analysis

Realizing the SQA vision requires translating the endoscopic video stream into a structured, semantic representation of the intraoperative scene. This translation is the central problem of computer vision in surgery: given a continuous flow of pixels, infer what is in the field, what is happening, and how it relates to the standards of the procedure. The information that AI can extract spans a hierarchy of increasingly abstract levels of scene understanding (Fig 2).

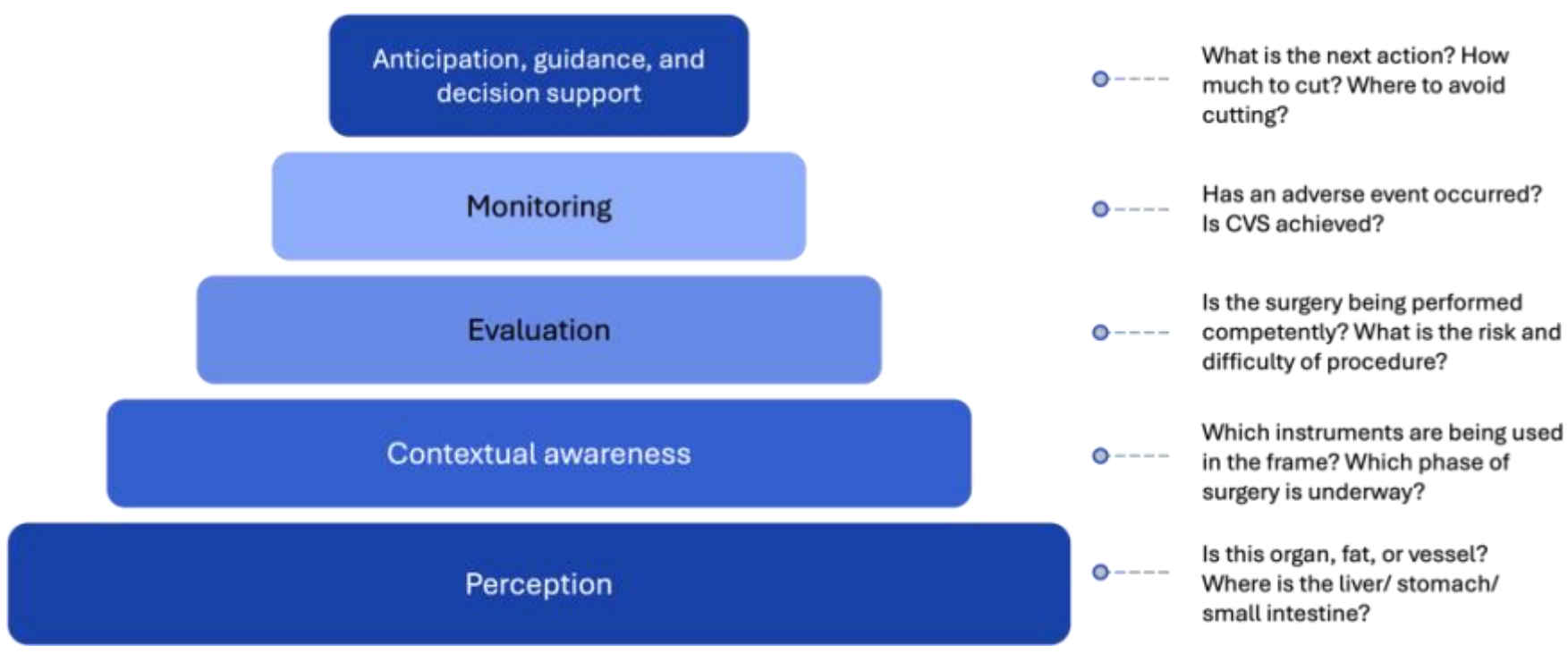

**Fig**. **2 Surgical scene understanding hierarchy**. Tasks span increasing levels of abstraction: from what is happening in the scene (perception), to where we are in the procedure (contextual awareness), to retrospective competency and skill assessment (evaluation), to early detection and real-time analysis (monitoring). Together these enable pre-emptive guidance and decision support, the most challenging end-goals of SQA.

At the most fundamental level, AI models are trained to recognize, localize, and delineate objects in the surgical scene, such as anatomical structures and surgical instruments. These capabilities, encompassing object detection, segmentation, and classification, form the building blocks upon which more complex analyses are constructed. For a model to learn such tasks, labelled data is required, where the objects of interest are annotated at varying levels of granularity, from image-level class labels to pixel-wise contour maps (Meireles et al. 2021). Addressing surgical safety, however, requires going beyond recognizing individual objects and developing a situational and contextual understanding of the scene. This involves modelling the spatial and temporal relationships between objects: how instruments interact with tissue, what action is being performed at a given moment, and how these actions unfold across the procedure. Surgical workflow analysis identifies the current phase or step of the operation, providing the situational awareness needed to monitor protocol adherence and detect deviations (Twinanda et al. 2017; Padoy 2019; Lavanchy et al. 2025). Surgical activity recognition models the instrument–tissue interaction at a finer granularity, often as triplets of instrument, verb, and target (Nwoye et al. 2022; Sharma et al. 2023). At a higher level still, models can be trained to anticipate upcoming events, recognize critical moments, or flag potentially unsafe situations, building the capabilities that underpin the safety applications described in the next section.

Translating these capabilities into systems that operate reliably in a real-world clinical environment requires a structured pipeline that extends well beyond training a model. The pipeline contains three broad stages: development, validation, and deployment. Each stage carries its own challenges and is where most projects either succeed or fail.

### *3.1 Development*

The development of a surgical AI system begins with data. Deep learning models are data-driven, and their performance depends on access to data that captures the variability of real practice: lighting conditions, camera angles, patient anatomy, instrument types, surgical techniques, and operator preferences. Acquiring such data in surgery is non-trivial, as it requires navigating ethical approvals, patient consent, and data protection regulations, in addition to the practical effort of recording, storing, and curating large volumes of high-resolution video (Mascagni et al. 2022).
Once data is acquired, the model must learn the target mapping from input to clinical concept. In supervised learning, this requires labelled examples annotated by experts. Acquiring these labels is typically the most resource-intensive part of the process; surgical annotation requires clinical expertise, is time-consuming, and is itself prone to inter-rater variability that must be controlled through clear annotation protocols and rater training (Ward et al. 2021; Mascagni et al. 2025). To reduce dependency on annotations, self-supervised approaches learn representations from the structure of the data itself, while semi-supervised and weakly-supervised approaches combine small, annotated sets with larger unlabeled or coarsely labelled corpora (Yengera et al. 2018; Vardazaryan et al. 2018; Nwoye et al. 2019). Foundation models pre-trained on large amounts of surgical video have begun to emerge and offer the prospect of transferring useful representations across procedures and tasks with substantially fewer annotations (Stilz et al., 2026).

The model itself is typically a deep neural network whose parameters are optimized by minimizing a loss function, representing a proxy for the target task such as the overlap between predicted and ground-truth segmentation masks or the accuracy of phase classification. The choice of loss function and evaluation metric shapes the behavior of the trained model and must reflect the clinical use case for which the system is intended.

### *3.2 Validation*

A model that performs well on its training data does not necessarily generalize to new patients, new centres, or new procedural contexts. Rigorous validation studies are therefore required before any AI system can be considered for clinical use. These involve evaluating models on held-out data not seen during training, ideally from institutions and patient populations that differ from those used for development. Multicenter studies subject the model to different data distributions and establish confidence in its performance under varying conditions (Mascagni et al. 2022; Bose et al. 2025).

Beyond aggregate performance metrics, a thorough validation must document failure cases and performance limitations, providing transparency about the conditions under which the model can and cannot be used reliably. For safety-critical applications, understanding when a model fails is as important as understanding when it succeeds. Prospective validation studies on real-world data and subjects form the basis for clinical evidence and are a often prerequisite for regulatory submission.

### *3.3 Deployment*

Deploying an AI system in the clinical environment introduces a further set of requirements. If the system has a clinical purpose, it is a medical device; no matter the embodiment (hardware or software) or its core technology (classical code or an AI/ML model). Approval from regulatory bodies, for example from the Food and Drug Administration (FDA) in the United States, is thus a prerequisite for clinical deployment. This entails rigorous documentation, risk analysis, evidence of safety and effectiveness, and provisions for post-market surveillance to monitor the system's performance once in use.

Operational constraints matter, too. The OR imposes hard limits on latency, computational budget, and physical footprint, and most ORs are not yet AI-ready in any technical sense. Recent work has shown that multiple computer vision models can be optimized and deployed concurrently on certified hardware to provide real-time analysis of laparoscopic video, with no malfunctions across early-stage clinical evaluations (Mascagni et al. 2024). Standards such as DICOM-RTV and IP-based streaming frameworks developed in the CONDOR project are addressing the data transmission side of the same problem (Mascagni and Padoy 2021).

Finally, design matters at least as much as AI performance, especially in high-stake environments such as the OR. For a successful operation, surgeons need to precisely manipulate instruments while making sense of the surgical scene; recalling patient-specific information and coordinating with other professionals in a timely manner. An AI system for intraoperative use needs to be carefully designed to deliver the right information, at the right time, to the right stakeholder. Mixed-methods studies following human factors and ergonomics principles can be used when designing and validating optimal human-machine interfaces such as copilots that aid surgical decision making.

Altogether, translating an AI system from a research prototype to a deployable clinical product is an arduous process that demands sustained interdisciplinary collaboration across clinical teams, data scientists, software engineers, designers, and regulatory affairs professionals (Mascagni et al. 2024).

## 4 Surgical AI Applications for Safety

Surgical safety, viewed through the SQA lens, is not a single quantity but a composite of interdependent factors: the complexity of the case, the proficiency of the surgeon, the compliance of the procedure to established standards, the moment-to-moment risk in the operative field, and the capacity to detect or pre-empt adverse events. AI-based analysis of endoscopic video, complemented where appropriate by audio and physiological signals from the wider OR, makes it possible to quantify each of these dimensions. The sections that follow describe how, organized by clinical concepts rather than by technical methods.

### *4.1 Quantitative assessment of surgical skill*

Surgery is a highly expert-driven discipline with a steep learning curve, and several studies establish a clear link between surgical skill and postoperative complications. The most influential is the bariatric surgery cohort of Birkmeyer and colleagues, which showed that surgeons whose technical skill rated in the bottom quartile of peer video review had higher rates of complications, reoperation, and death than those in the top quartile (Birkmeyer et al. 2013). Skill assessment therefore matters not only for training but for outcomes. Traditionally it has relied on direct observation by expert surgeons, in real time or via retrospective video review, scored on rubrics such as the Objective Structured Assessment of Technical Skills (OSATS), the Global Evaluative Assessment of Robotic Skills (GEARS), or procedure-specific CATs. Manual assessment is laborious, prone to inter-rater variability, and constrained by the availability of expert assessors.

AI-driven methods offer a path to automated, scalable, and objective skill assessment by learning the relationship between visual features of the surgical video and corresponding skill ratings (Chen et al. 2024). Models are trained on labelled examples of videos or video segments scored by expert assessors and then learn to predict skill levels from the video alone. Convolutional and recurrent architectures, and more recently transformer-based models, can distinguish novice, intermediate, and expert performance by extracting spatial and temporal features from the stream (Lavanchy et al. 2021). Phase-recognition models trained for workflow analysis have been repurposed for skill assessment, since the temporal dynamics of phase transitions, time spent in each phase, and the order of steps correlate with proficiency (Anastasiou et al. 2025). A related line of work links automatically extracted tool-motion features with technical skill scores (Jin et al. 2018; Lavanchy et al. 2021), and recent methods explore few-shot and self-supervised approaches to reduce annotation requirements (Anastasiou et al. 2025). These methods do not yet replace human raters, but they are beginning to supplement them, and at sufficient maturity they could turn skill assessment from a periodic to a continuous practice (Sestini et al., 2025).

### *4.2 Surgical risk and quality measures*

Surgical quality encompasses a broader set of considerations than competence alone, addressing whether the procedure progresses in accordance with established safety standards. The CVS for laparoscopic cholecystectomy has been the most active testing ground for AI-based quality measurement. Several deep learning models have been developed to segment the hepatocystic anatomy and predict whether each CVS criterion is met from endoscopic images. Mascagni and colleagues introduced DeepCVS, a two-stage model combining segmentation and classification that achieved performance approaching that of expert raters (Mascagni et al. 2022b). Public datasets such as Cholec80, Cholec80-CVS, and Endoscapes have accelerated this work by providing annotated video for model development and benchmarking (Twinanda et al. 2017; Mascagni et al. 2025).In addition, large international initiatives such as the SAGES Critical View of Safety Challenge are gathering more than 1,000 videos with clinically meaningful annotations (Mascagni et al. 2024).

A complementary approach delineates safe and dangerous zones within the operative field. Madani and colleagues developed GoNoGoNet, trained on expert annotations of safe and unsafe areas of dissection in laparoscopic cholecystectomy, providing a color-coded overlay that flags when the surgeon approaches a region of elevated risk (Madani et al. 2022). Similar models have been built to identify anatomical planes of dissection in colorectal and hepato-pancreato-biliary procedures, providing a form of intraoperative navigation that does not require external tracking or preoperative imaging registration (Kolbinger et al. 2023; Ryu et al. 2024). Other models target specific intraoperative actions, such as ClipAssistNet for clip placement (Aspart et al. 2022) and detectors for bleeding events. Together, the CVS, go/no-go zones, plane detection, and action-specific assistants form a hierarchy of quality measures, from discrete checkpoints through spatial risk maps to continuous anatomical guidance, that AI systems can monitor and surface in real time.

Beyond gallbladder surgery, computer vision is being applied to a widening range of procedures and quality questions: assessment of gallbladder inflammation as a surrogate of operative difficulty (Ward et al. 2022; Loukas et al. 2021), automatic surgical reporting (Berlet et al. 2022), and condensed video documentation of critical events such as the EndoDigest platform, which has been validated in a multicentric study (Mascagni et al. 2021b, 2022d).

### *4.3 Detecting intraoperative adverse events*

Comprehensive SQA requires moving beyond compliance with explicit standards to recognition of departures from safe practice and detection of adverse events. Intraoperative adverse events occur in approximately 2% of operations (Mavros et al. 2014), span a wide severity range (Jung and Grantcharov 2019), and are widely under-reported in conventional records, hindering the institutional learning that quality improvement depends on (Goldenberg et al. 2017; Curtis et al. 2021).

Detecting adverse events automatically from endoscopic video is technically harder than recognizing instruments or anatomy. Adverse events arise from a complex interplay of anatomical, technical, and situational factors and do not present with a consistent visual signature. Detecting them requires integrating multiple contextual cues, such as tissue appearance, instrument behavior, surgical phase, and deviations from procedural patterns. Adverse events are also rare relative to routine activity, producing highly imbalanced datasets in which models must learn patterns from a small fraction of the training data (Bose et al. 2025). Their temporal dynamics pose a further challenge, as the boundary between precursors, events, and consequence is often ambiguous, requiring models that reason over extended temporal windows. Recent work has shown that mixing-style architectures and triplet-based action recognition can begin to detect classes of intraoperative adverse events, such as bleeding, in laparoscopic Roux-en-Y gastric bypass (Lavanchy et al. 2025; Bose et al. 2025).

Two related concepts deserve mention. The first is the near miss: an event in which an adverse outcome was narrowly avoided. Near misses are by definition absent from conventional records and can only be captured through continuous video-based monitoring, where their value for quality improvement is large precisely because they reveal failure modes without harm to patients (Curtis et al. 2021). The second is the analysis of OR-level rather than surgical-field-level adverse events. AI-based audiovisual analysis can monitor team communication, compliance with standard operating procedures, and the prevalence of distractions, complementing the surgical-field analyses described above (Goldenberg et al. 2017; Peisl et al. 2024). Both lines of work raise important considerations around privacy, consent, and OR culture that must be navigated carefully (Lavanchy et al. 2023; De Backer et al. 2024).

### *4.4 Anticipating critical events*

Anticipating critical events to enable preventive action represents the next frontier and the most technically demanding direction. The central difficulty is robust temporal modelling of the surgical procedure: the AI system must learn to map patterns in the video stream to predicted events with sufficient lead time. Architectures capable of reasoning over extended sequences of frames, including recurrent networks, temporal convolutional networks, and increasingly transformer-based models, are well suited to this task. The anticipation horizon imposes an inherent trade-off: a longer prediction window provides more time for intervention but demands extrapolation from increasingly ambiguous cues, while a shorter window may offer greater specificity at the cost of clinical utility.

Early work has demonstrated that anticipation is feasible in principle. RSDNet predicts remaining surgery duration from laparoscopic video without manual annotations and could inform OR scheduling and team readiness (Twinanda et al. 2019). Models for early surgery type recognition can identify the procedure being performed from the first minutes of video (Kannan et al. 2020). Multimodal approaches that combine endoscopic video with vital signs, robotic kinematics, and procedural metadata are beginning to extend the horizon and improve specificity. These results remain confined to retrospective evaluations on curated datasets, and the path from proof-of-concept to clinically deployed anticipation systems will require improvements in model robustness as well as careful design of how alerts are integrated into the surgical workflow.

## 5 The Road Ahead

To borrow a useful analogy, surgery is converging on something it has long lacked: a layer of continuous, sensor-driven situational awareness comparable to what aviation built around flight recorders, on-board instrumentation, and air traffic control. The aviation parallel is informative for the safety culture and decision-support architectures it offers, but the analogy has limits; the state of the surgical field, with its complex, deformable anatomy and high inter-patient variability, is fundamentally less measurable than the state of an aircraft. The works presented in the previous sections demonstrate that the technical foundations for AI-enabled SQA exist, but we need to move from proof-of-concept to clinical value. Closing the gap from demonstration to value requires sustained progress across data, methods, regulation, and governance.

### *5.1 Data: from siloed videos to representative, shared resources*

Curated, large-scale surgical data representative of the real-world, is the foundation on which everything else rests. Existing datasets are concentrated in a small number of academic centres in high-income countries, biasing the technologies built on them. Even where centres are willing to share, current clinical infrastructure is poorly equipped for high-volume video storage, and the medico-legal frameworks governing data sharing are unclear to say the least. Fortunately, a multi-pronged response is taking shape. Federated learning enables model training across institutions without centralizing raw data, mitigating privacy and sovereignty concerns (Kassem et al. 2023). Publicly released datasets and challenges, including Cholec80, Endoscapes, and the SAGES CVS Challenge, set common benchmarks and lower the barrier of entry for new groups (Twinanda et al. 2017; Mascagni et al. 2025). Privacy-preserving pipelines for endoscopic videos, such as out-of-body classifiers or surgeon-friendly applications for video de-identification such as EndoShare, address the practical obstacles to sharing (Lavanchy et al. 2023; Arboit et al. 2025). Behavioral and statutory levers are equally important: institutions that combine clear policies with concrete incentives, such as continuing medical education credits and patient safety organization protections, succeed in capturing more video and more useful video than those that do not (Mazer et al. 2022; Mascagni et al. 2022).

### *5.2 Methods: from accuracy to robustness, explainability, and modular design*

Even with sufficient data, a model that achieves high accuracy on a curated benchmark may fail in real-world surgical practice. Three methodological priorities are emerging. The first is explainability, not as a research curiosity but as a precondition for understanding failure modes and contraindications of AI systems; without it, regulatory and clinicians trust is hard to earn (Mascagni et al. 2024b). The second is robustness, the ability of a system to reliably work across settings (Maier-Hein et al. 2022). The third is a modular, hierarchical design, in which complex tasks are decomposed into simpler ones whose outputs can be combined and inspected. EndoDigest is a worked example, combining a deep workflow recognizer with a rule-based event localizer to produce condensed video documentation that performs well even when individual components are imperfect (Mascagni et al. 2021b, 2022d). Modular designs are easier to validate, easier to audit, and easier to repurpose across procedures.

### *5.3 Evaluation: from technical metrics to clinical value*

Technical performance and regulatory clearance do not by themselves guarantee clinical adoption or patient benefit. Prospective, multi-center clinical studies that evaluate workflow integration, surgeon and team experience, and patient outcomes are essential to build the evidence base for routine use. AI is not a pill, and trial designs imported wholesale from pharmacology will not always fit. Most surgical

AI systems act on how care is delivered rather than on the patient directly, which raises the risk of contamination bias when the same surgeon participates in both arms of a randomized trial. Cluster-randomized designs and stepped-wedge trials, which randomize at the level of surgeon or hospital and include both before–after and parallel-control elements, are better suited (McClure et al. 2020; Hemming et al. 2015). Pragmatic real-world studies in adjacent fields have already shown that AI assistance can be used differently from how its developers intended, sometimes to neutral or negative effect, underscoring the need for outcome-anchored evaluation rather than technical-metric evaluation alone (Levy et al. 2022; Ladabaum et al. 2023; van Leeuwen et al. 2024).

## *5.4 Governance: regulation, liability, and equitable access*

As discussed, Software as a Medical Devices requires a defined intended use and evidence of safety and efficacy proportionate to the risk class. Regulators face a moving target, and intended use as documented does not always match how a product is used in practice (van Leeuwen et al. 2024). Constructive collaboration between regulators, manufacturers, and clinical communities is needed to establish efficient and transparent pathways. Liability remains an open question; current AI systems estimate information rather than make autonomous decisions, so their introduction into surgical practice is best framed as a new class of medical device with established risk-management principles, but the pace of capability change will continue to test this framing (Cobianchi et al. 2022; Mascagni et al. 2024).
Furthermore, equitable access deserves explicit attention. Datasets, validation studies, and commercial deployments are heavily concentrated in high-income, high-resource settings, and the same forces that propelled MIS in those settings are not equally distributed worldwide. Without active counter-effort, AI risks widening rather than narrowing surgical disparities (Knight et al. 2019; Meara et al. 2015). Multi-center and global collaborative initiatives, such as the SAGES CVS Challenge, illustrate what is possible when academic, clinical, and industrial partners coordinate around shared infrastructure (Alapatt et al. 2025).

## *5.5 From a fragmented practice to a learning surgical system*

Surgery has seen three major paradigm shifts: general anesthesia, asepsis, and minimally invasive surgery. The combination of comprehensive sensing, AI-driven analysis, and closed-loop assistance is positioning the field for a fourth, characterized by smart augmentation of perceptual, cognitive, and physical tasks (Maier-Hein et al. 2017; Mascagni and Padoy 2021). The trajectory is likely to be gradual rather than disruptive, and rightly so: surgical safety is too consequential to be hurried by hype or held back by inertia. The works described in this chapter, from the CVS to CATs to real-time AI assistance, all share an underlying intent: to turn surgery from a practice that is studied indirectly through outcomes into one that is observed, measured, and improved continuously. AI-enabled Surgical Quality Assurance is the operational expression of that intent, and its maturation over the coming years will depend less on any single algorithm than on the quality of the data, methods, evidence, and governance that the surgical community builds together.

## *Acknowledgments*

This work was supported by the European Union (ERC, CompSURG, 101088553). Views and opinions expressed are however those of the authors only and do not necessarily reflect those of the European Union or the European Research Council. Neither the European Union nor the granting authority can be held responsible for them. This work has also been supported by French state funds managed within the Plan Investissements d'Avenir by the ANR under reference ANR- 10-IAHU-02 (IHU Strasbourg).

## *Disclosures*

PM, DA, NP are co-founders and shareholders of Scialytics. LS does not have any conflict of interests to disclose.

# References

Alapatt, D., Eckhoff, J., Lyu, Z., Ban, Y., Mazellier, J.-P., Choksi, S., Yang, K., Chiang, P.-H., Zorzetti, N., Cannas, S., Neimark, D., Bar, O., Yamlahi, A., Hennighausen, J., Wang, X., Li, R., Liang, L., Wang, Y., Koju, S., … Padoy, N. (2025). The SAGES Critical View of Safety Challenge: A Global Benchmark for AI-Assisted Surgical Quality Assessment (Version 2). arXiv. https://doi.org/10.48550/ARXIV.2509.17100

Anastasiou D, Caramalau R, Sirajudeen N, et al. (2025). Exploring Pre-Training Across Domains for Few-Shot Surgical Skill Assessment. arXiv:2509.09327. https://doi.org/10.48550/arXiv.2509.09327

Arboit, L., Schneider, D. N., Baby, B., Srivastav, V., Mascagni, P., & Padoy, N. (2025). Endoshare: A Publicly Available, Surgeons-Friendly Solution to De-Identify and Manage Surgical Videos (Version 2). arXiv. https://doi.org/10.48550/ARXIV.2510.20087

Aspart F, Bolmgren JL, Lavanchy JL, et al. (2022). ClipAssistNet: bringing real-time safety feedback to operating rooms. International Journal of Computer Assisted Radiology and Surgery 17(1): 5–13. https://doi.org/10.1007/s11548-021-02441-x

Berci G, Hunter J, Morgenstern L, et al. (2013). Laparoscopic cholecystectomy: first, do no harm; second, take care of bile duct stones. Surgical Endoscopy 27(4): 1051–1054. https://doi.org/10.1007/s00464-012-2767-5

Berlet M, Vogel T, Ostler D, et al. (2022). Surgical reporting for laparoscopic cholecystectomy based on phase annotation by a convolutional neural network (CNN) and the phenomenon of phase flickering: a proof of concept. International Journal of Computer Assisted Radiology and Surgery. https://doi.org/10.1007/s11548-022-02680-6

Birkmeyer JD, Finks JF, O'Reilly A, et al. (2013). Surgical Skill and Complication Rates after Bariatric Surgery. New England Journal of Medicine 369(15): 1434–1442. https://doi.org/10.1056/NEJMsa1300625

Bose R, Nwoye CI, Lazo J, Lavanchy JL, Padoy N. (2025). Feature Mixing Approach for Detecting Intraoperative Adverse Events in Laparoscopic Roux-En-Y Gastric Bypass Surgery. arXiv:2504.16749. https://doi.org/10.48550/arXiv.2504.16749

Brunt LM, Deziel DJ, Telem DA, et al. (2020). Safe Cholecystectomy Multi-society Practice Guideline and State of the Art Consensus Conference on Prevention of Bile Duct Injury During Cholecystectomy. Annals of Surgery 272(1): 3–23. https://doi.org/10.1097/SLA.0000000000003791

Chen K, Schewski L, Srivastav V, et al. (2025a). When do they StOP?: A First Step Towards Automatically Identifying Team Communication in the Operating Room. arXiv:2502.08299. https://doi.org/10.48550/arXiv.2502.08299

Chen, K., Srivastav, V., Mutter, D., & Padoy, N. (2025b). Learning from Synchronization: Self-Supervised Uncalibrated Multi-View Person Association in Challenging Scenes. https://arxiv.org/abs/2503.13739

Chen Z, Yang D, Li A, et al. (2024). Decoding Surgical Skill: An Objective and Efficient Algorithm for Surgical Skill Classification Based on Surgical Gesture Features – Experimental Studies. International Journal of Surgery 110(3): 1441–1449. https://doi.org/10.1097/JS9.0000000000000975

Cobianchi L, Verde JM, Loftus TJ, et al. (2022). Artificial Intelligence and Surgery: Ethical Dilemmas and Open Issues. Journal of the American College of Surgeons 235(2): 268–275. https://doi.org/10.1097/XCS.0000000000000242

Conley DM, Singer SJ, Edmondson L, Berry WR, Gawande AA. (2011). Effective Surgical Safety Checklist Implementation. Journal of the American College of Surgeons 212(5): 873–879. https://doi.org/10.1016/j.jamcollsurg.2011.01.052

Curtis NJ, Dennison G, Brown CSB, et al. (2021). Clinical Evaluation of Intraoperative Near Misses in Laparoscopic Rectal Cancer Surgery. Annals of Surgery 273(4): 778–784. https://doi.org/10.1097/SLA.0000000000003452

Daes, J., & Felix, E. (2017). Critical View of the Myopectineal Orifice. Annals of surgery, 266(1), e1–e2. https://doi.org/10.1097/SLA.0000000000002104

Degiuli, M., Reddavid, R., Tomatis, M., Ponti, A., Morino, M., Sasako, M., & of the Italian Gastric Cancer Study Group (IGCSG) (2021). D2 dissection improves disease-specific survival in advanced gastric cancer patients: 15-year follow-up results of the Italian Gastric Cancer Study Group D1 versus D2 randomised controlled trial. European journal of cancer (Oxford, England : 1990), 150, 10–22. https://doi.org/10.1016/j.ejca.2021.03.031

De Backer P, Simoens J, Mestdagh K, et al. (2024). Privacy-proof live surgery streaming: development and validation of a low-cost, real-time robotic surgery anonymization algorithm. Annals of Surgery. https://doi.org/10.1097/SLA.0000000000006245

Diamond IR, Grant RC, Feldman BM, et al. (2014). Defining consensus: a systematic review recommends methodologic criteria for reporting of Delphi studies. Journal of Clinical Epidemiology 67(4): 401–409. https://doi.org/10.1016/j.jclinepi.2013.12.002

Dias RD, Kennedy-Metz LR, Yule SJ, Gombolay M, Zenati MA. (2022). Assessing Team Situational Awareness in the Operating Room via Computer Vision. IEEE CogSIMA 2022: 94–96.

Fowler AJ, Wan YI, Prowle JR, et al. (2022). Long-term mortality following complications after elective surgery: a secondary analysis of pooled data from two prospective cohort studies. British Journal of Anaesthesia 129(4): 588–597. https://doi.org/10.1016/j.bja.2022.06.019

Fuchs, B., and P. Heesen. (2025). 'From Lindbergh's Cockpit to Predictive, AI-Driven Health Care: A Roadmap for High-Reliability Medicine'. ESMO Real World Data and Digital Oncology 10 (December): 100650. https://doi.org/10.1016/j.esmorw.2025.100650.

Gogalniceanu P, Calder F, Callaghan C, Sevdalis N, Mamode N. (2021). Surgeons Are Not Pilots: Is the Aviation Safety Paradigm Relevant to Modern Surgical Practice? Journal of Surgical Education 78(5): 1393–1399. https://doi.org/10.1016/j.jsurg.2021.01.016

Goldenberg MG, Jung J, Grantcharov TP. (2017). Using Data to Enhance Performance and Improve Quality and Safety in Surgery. JAMA Surgery 152(10): 972–973. https://doi.org/10.1001/jamasurg.2017.2888

Hamoud, I., Jamal, M.A., Srivastav, V., MUTTER, D., Padoy, N. & Mohareri, O.. (2024). ST(OR)$^2$: Spatio-Temporal Object Level Reasoning for Activity Recognition in the Operating Room. Medical Imaging with Deep Learning, in Proceedings of Machine Learning Research 227:1254-1268 https://proceedings.mlr.press/v227/hamoud24a.html.

Haynes AB, Weiser TG, Berry WR, et al. (2009). A Surgical Safety Checklist to Reduce Morbidity and Mortality in a Global Population. New England Journal of Medicine 360(5): 491–499. https://doi.org/10.1056/NEJMsa0810119

Healy MA, Mullard AJ, Campbell DA Jr, Dimick JB. (2016). Hospital and Payer Costs Associated with Surgical Complications. JAMA Surgery 151(9): 823–830. https://doi.org/10.1001/jamasurg.2016.0773

Helmreich RL. (2000). On error management: lessons from aviation. BMJ 320(7237): 781–785. https://doi.org/10.1136/bmj.320.7237.781

Hemming K, Haines TP, Chilton PJ, Girling AJ, Lilford RJ. (2015). The stepped wedge cluster randomised trial: rationale, design, analysis, and reporting. BMJ 350: h391. https://doi.org/10.1136/bmj.h391

International Surgical Outcomes Study Group. (2016). Global patient outcomes after elective surgery: prospective cohort study in 27 low-, middle- and high-income countries. British Journal of Anaesthesia 117(5): 601–609. https://doi.org/10.1093/bja/aew316

Jin A, Yeung S, Jopling J, et al. (2018). Tool Detection and Operative Skill Assessment in Surgical Videos Using Region-Based Convolutional Neural Networks. IEEE Winter Conference on Applications of Computer Vision (WACV): 691–699. https://doi.org/10.1109/WACV.2018.00081

Jung JJ, Grantcharov TP. (2019). Development and Evaluation of a Novel Instrument to Measure Severity of Intraoperative Events Using Video Data. Journal of the American College of Surgeons 229(4, Suppl 1): S148–S149. https://doi.org/10.1016/j.jamcollsurg.2019.08.328

Jung, J. J., Elfassy, J., & Grantcharov, T. (2020). Factors associated with surgeon's perception of distraction in the operating room. Surgical endoscopy, 34(7), 3169–3175. https://doi.org/10.1007/s00464-019-07088-z

Kannan S, Yengera G, Mutter D, Marescaux J, Padoy N. (2020). Future-State Predicting LSTM for Early Surgery Type Recognition. IEEE Transactions on Medical Imaging 39(3): 556–566. https://doi.org/10.1109/TMI.2019.2931158

Kassem H, Alapatt D, Mascagni P, Karargyris A, Padoy N. (2023). Federated Cycling (FedCy): Semi-Supervised Federated Learning of Surgical Phases. IEEE Transactions on Medical Imaging 42(7): 1920–1931. https://doi.org/10.1109/TMI.2022.3222126

Kennedy-Metz LR, Dias RD, Srey R, et al. (2020). Sensors for Continuous Monitoring of Surgeon's Cognitive Workload in the Cardiac Operating Room. Sensors 20(22): 6616. https://doi.org/10.3390/s20226616

Knight SR, Ots R, Maimbo M, et al. (2019). Systematic review of the use of big data to improve surgery in low- and middle-income countries. British Journal of Surgery 106(2): e62–e72. https://doi.org/10.1002/bjs.11052

Kolbinger FR, Bodenstedt S, Carstens M, et al. (2023). Artificial Intelligence for context-aware surgical guidance in complex robot-assisted oncological procedures: an exploratory feasibility study. European Journal of Surgical Oncology: 106996. https://doi.org/10.1016/j.ejso.2023.106996

Korndorffer, J. R., Jr, Hawn, M. T., Spain, D. A., Knowlton, L. M., Azagury, D. E., Nassar, A. K., Lau, J. N., Arnow, K. D., Trickey, A. W., & Pugh, C. M. (2020). Situating Artificial Intelligence in Surgery: A Focus on Disease Severity. Annals of surgery, 272(3), 523–528. https://doi.org/10.1097/SLA.0000000000004207

Ladabaum U, Shepard J, Weng Y, et al. (2023). Computer-aided detection of polyps does not improve colonoscopist performance in a pragmatic implementation trial. Gastroenterology 164(3): 481–483.e6. https://doi.org/10.1053/j.gastro.2022.12.004

Lavanchy JL, Alapatt D, Sestini L, et al. (2025). Analyzing the Impact of Surgical Technique on Intraoperative Adverse Events in Laparoscopic Roux-En-Y Gastric Bypass Surgery by Video-Based Assessment. Surgical Endoscopy 39(3): 2026–2036. https://doi.org/10.1007/s00464-025-11557-z

Lavanchy JL, Vardazaryan A, Mascagni P, Mutter D, Padoy N. (2023). Preserving privacy in surgical video analysis using a deep learning classifier to identify out-of-body scenes in endoscopic videos. Scientific Reports 13: 9235. https://doi.org/10.1038/s41598-023-36453-1

Lavanchy JL, Zindel J, Kirtac K, et al. (2021). Automation of surgical skill assessment using a three-stage machine learning algorithm. Scientific Reports 11: 5197. https://doi.org/10.1038/s41598-021-84295-6

Levy I, Bruckmayer L, Klang E, Ben-Horin S, Kopylov U. (2022). Artificial intelligence-aided colonoscopy does not increase adenoma detection rate in routine clinical practice. American Journal of Gastroenterology 117(11): 1871–1873. https://doi.org/10.14309/ajg.0000000000001970

Loukas C, Frountzas M, Schizas D. (2021). Patch-based classification of gallbladder wall vascularity from laparoscopic images using deep learning. International Journal of Computer Assisted Radiology and Surgery 16(1): 103–113. https://doi.org/10.1007/s11548-020-02285-x

Madani A, Namazi B, Altieri MS, et al. (2022). Artificial Intelligence for Intraoperative Guidance: Using Semantic Segmentation to Identify Surgical Anatomy During Laparoscopic Cholecystectomy. Annals of Surgery 276(2): 363–369. https://doi.org/10.1097/SLA.0000000000004594

Maier-Hein L, Eisenmann M, Sarikaya D, et al. (2022). Surgical data science – from concepts toward clinical translation. Medical Image Analysis 76: 102306. https://doi.org/10.1016/j.media.2021.102306

Maier-Hein L, Vedula SS, Speidel S, et al. (2017). Surgical data science for next-generation interventions. Nature Biomedical Engineering 1: 691–696. https://doi.org/10.1038/s41551-017-0132-7

Mascagni P, Alapatt D, Lapergola A, et al. (2024b). Early-Stage Clinical Evaluation of Real-Time Artificial Intelligence Assistance for Laparoscopic Cholecystectomy. British Journal of Surgery 111(1): znad353. https://doi.org/10.1093/bjs/znad353

Mascagni P, Alapatt D, Laracca GG, et al. (2022c). Multicentric validation of EndoDigest: a computer vision platform for video documentation of the critical view of safety in laparoscopic cholecystectomy. Surgical Endoscopy 36(11): 8379–8386. https://doi.org/10.1007/s00464-022-09112-1

Mascagni P, Alapatt D, Murali A, et al. (2025). Endoscapes, a Critical View of Safety and Surgical Scene Segmentation Dataset for Laparoscopic Cholecystectomy. Scientific Data 12(1): 331. https://doi.org/10.1038/s41597-025-04642-4
Mascagni P, Alapatt D, Sestini L, et al. (2022). Computer vision in surgery: from potential to clinical value. npj Digital Medicine 5: 163. https://doi.org/10.1038/s41746-022-00707-5
Mascagni P, Alapatt D, Sestini L, et al. (2024). Applications of artificial intelligence in surgery: clinical, technical, and governance considerations. Cirugía Española 102(S1): 66–71. https://doi.org/10.1016/j.cireng.2024.04.009
Mascagni P, Alapatt D, Urade T, et al. (2021b). A Computer Vision Platform to Automatically Locate Critical Events in Surgical Videos: Documenting Safety in Laparoscopic Cholecystectomy. Annals of Surgery 274(1): e93–e95. https://doi.org/10.1097/SLA.0000000000004736
Mascagni P, Alapatt D, Urade T, et al. (2022d). Response to comments on: A Computer Vision Platform to Automatically Locate Critical Events in Surgical Videos: Documenting Safety in Laparoscopic Cholecystectomy. Annals of Surgery 276(6): e637. https://doi.org/10.1097/SLA.0000000000005267
Mascagni P, Padoy N. (2021). OR black box and surgical control tower: Recording and streaming data and analytics to improve surgical care. Journal of Visceral Surgery 158(3S): S18–S25. https://doi.org/10.1016/j.jviscsurg.2021.01.004
Mascagni P, Rodríguez-Luna MR, Urade T, et al. (2021). Intraoperative Time-out to Promote the Implementation of the Critical View of Safety in Laparoscopic Cholecystectomy: A Video-Based Assessment of 343 Procedures. Journal of the American College of Surgeons 233(4): 497–505. https://doi.org/10.1016/j.jamcollsurg.2021.06.018
Mascagni P, Vardazaryan A, Alapatt D, et al. (2022b). Artificial Intelligence for Surgical Safety: Automatic Assessment of the Critical View of Safety in Laparoscopic Cholecystectomy Using Deep Learning. Annals of Surgery 275(5): 955–961. https://doi.org/10.1097/SLA.0000000000004351
Mavros MN, Velmahos GC, Larentzakis A, et al. (2014). Opening Pandora's box: understanding the nature, patterns, and 30-day outcomes of intraoperative adverse events. American Journal of Surgery 208(4): 626–631. https://doi.org/10.1016/j.amjsurg.2014.02.014
Mazer L, Varban O, Montgomery JR, Awad MM, Schulman A. (2022). Video is better: why aren't we using it? A mixed-methods study of the barriers to routine procedural video recording and case review. Surgical Endoscopy 36(2): 1090–1097. https://doi.org/10.1007/s00464-021-08375-4
McClure GR, Belley-Côté EP, Spence J, Lee SF, Whitlock RP. (2020). Selecting the optimal level of clustering: an approach to trial design decision making. Journal of the American College of Surgeons 231(3): 397. https://doi.org/10.1016/j.jamcollsurg.2020.05.012
McCulloch P, Altman DG, Campbell WB, et al. (2009). No surgical innovation without evaluation: the IDEAL recommendations. Lancet 374(9695): 1105–1112. https://doi.org/10.1016/S0140-6736(09)61116-8
Meara JG, Leather AJM, Hagander L, et al. (2015). Global Surgery 2030: Evidence and solutions for achieving health, welfare, and economic development. Surgery 158(1): 3–6. https://doi.org/10.1016/j.surg.2015.04.011
Meireles OR, Rosman G, Altieri MS, et al. (2021). SAGES consensus recommendations on an annotation framework for surgical video. Surgical Endoscopy 35(9): 4918–4929. https://doi.org/10.1007/s00464-021-08578-9
Miskovic D, Ni M, Wyles SM, et al. (2013). Is competency assessment at the specialist level achievable? A study for the national training programme in laparoscopic colorectal surgery in England. Annals of Surgery 257(3): 476–482. https://doi.org/10.1097/SLA.0b013e318275b72a
Nwoye CI, Mutter D, Marescaux J, Padoy N. (2019). Weakly supervised convolutional LSTM approach for tool tracking in laparoscopic videos. International Journal of Computer Assisted Radiology and Surgery 14(6): 1059–1067. https://doi.org/10.1007/s11548-019-01958-6

Nwoye CI, Yu T, Gonzalez C, et al. (2022). Rendezvous: Attention mechanisms for the recognition of surgical action triplets in endoscopic videos. Medical Image Analysis 78: 102433. https://doi.org/10.1016/j.media.2022.102433

Padoy N. (2019). Machine and deep learning for workflow recognition during surgery. Minimally Invasive Therapy & Allied Technologies 28(2): 82–90. https://doi.org/10.1080/13645706.2019.1584116

Peisl S, Sánchez-Taltavull D, Guillen-Ramirez H, et al. (2024). Noise in the operating room coincides with surgical difficulty. BJS Open 8(5): zrae098. https://doi.org/10.1093/bjsopen/zrae098

Rex DK, Schoenfeld PS, Cohen J, et al. (2015). Quality indicators for colonoscopy. American Journal of Gastroenterology 110(1): 72–90. https://doi.org/10.1038/ajg.2014.385

Ryu S, Goto K, Imaizumi Y, Nakabayashi Y. (2024). Laparoscopic Colorectal Surgery with Anatomical Recognition with Artificial Intelligence Assistance for Nerves and Dissection Layers. Annals of Surgical Oncology 31: 1690–1691. https://doi.org/10.1245/s10434-023-14633-7

Sestini, L., Harris, M., Alapatt, D., Yu, T., Mascagni, P., Swanström, L. L., Padoy, N., & Shlomovitz, E. (2025). Artificial intelligence for automatic FLS credentialing: Highlighting and addressing current limitations. Surgical Endoscopy, 39(10), 7028–7034. https://doi.org/10.1007/s00464-025-12076-7

Sharma S, Nwoye CI, Mutter D, Padoy N. (2023). Surgical Action Triplet Detection by Mixed Supervised Learning of Instrument-Tissue Interactions. In: MICCAI 2023. Springer Nature Switzerland. https://doi.org/10.1007/978-3-031-43996-4_48

Stilz, F., Srivastav, V., Navab, N., & Padoy, N. (2026). CliPPER: Contextual Video-Language Pretraining on Long-form Intraoperative Surgical Procedures for Event Recognition (Version 1). arXiv. https://doi.org/10.48550/ARXIV.2603.24539

Strasberg SM, Hertl M, Soper NJ. (1995). An analysis of the problem of biliary injury during laparoscopic cholecystectomy. Journal of the American College of Surgeons 180(1): 101–125. PMID: 8000648

Subramanian, A., & Liang, M. K. (2012). A 60-year literature review of stump appendicitis: The need for a critical view. The American Journal of Surgery, 203(4), 503–507. https://doi.org/10.1016/j.amjsurg.2011.04.009

Sullivan ME, Brown CVR, Peyre SE, et al. (2008). The use of cognitive task analysis to improve the learning of percutaneous tracheostomy placement. American Journal of Surgery 195(2): 167–171. https://doi.org/10.1016/j.amjsurg.2006.09.005

Treadwell JR, Lucas S, Tsou AY. (2014). Surgical checklists: a systematic review of impacts and implementation. BMJ Quality & Safety 23(4): 299–318. https://doi.org/10.1136/bmjqs-2012-001797

Tsai AY-C, Mavroveli S, Miskovic D, et al. (2019). Surgical Quality Assurance in COLOR III: Standardization and Competency Assessment in a Randomized Controlled Trial. Annals of Surgery 270(5): 768–774. https://doi.org/10.1097/SLA.0000000000003537

Tschan F, Keller S, Semmer NK, et al. (2022). Effects of structured intraoperative briefings on patient outcomes: multicentre before-and-after study. British Journal of Surgery 109(1): 136–144. https://doi.org/10.1093/bjs/znab384

Twinanda AP, Shehata S, Mutter D, Marescaux J, de Mathelin M, Padoy N. (2017). EndoNet: A Deep Architecture for Recognition Tasks on Laparoscopic Videos. IEEE Transactions on Medical Imaging 36(1): 86–97. https://doi.org/10.1109/TMI.2016.2593957

Twinanda AP, Yengera G, Mutter D, Marescaux J, Padoy N. (2019). RSDNet: Learning to Predict Remaining Surgery Duration from Laparoscopic Videos Without Manual Annotations. IEEE Transactions on Medical Imaging 38(4): 1069–1078. https://doi.org/10.1109/TMI.2018.2878055

Vardazaryan A, Mutter D, Marescaux J, Padoy N. (2018). Weakly-Supervised Learning for Tool Localization in Laparoscopic Videos. In: Intravascular Imaging and Computer Assisted Stenting and Large-Scale Annotation of Biomedical Data and Expert Label Synthesis. Springer. https://doi.org/10.1007/978-3-030-01364-6_19

Vasey B, Nagendran M, Campbell B, et al. (2022). Reporting guideline for the early-stage clinical evaluation of decision support systems driven by artificial intelligence: DECIDE-AI. Nature Medicine 28: 924–933. https://doi.org/10.1038/s41591-022-01772-9

Ward TM, Hashimoto DA, Ban Y, Rosman G, Meireles OR. (2022). Artificial intelligence prediction of cholecystectomy operative course from automated identification of gallbladder inflammation. Surgical Endoscopy 36(9): 6832–6840. https://doi.org/10.1007/s00464-022-09009-z

Ward, T. M., Fer, D. M., Ban, Y., Rosman, G., Meireles, O. R., & Hashimoto, D. A. (2021). Challenges in surgical video annotation. Computer Assisted Surgery, 26(1), 58–68. https://doi.org/10.1080/24699322.2021.1937320

Watson SL, Fowler AJ, Pearse RM, Abbott TEF. (2025). The financial cost of postoperative complications: a prospective cohort study with linked health systems data. British Journal of Anaesthesia 135: 1586–1590. https://doi.org/10.1016/j.bja.2025.06.039

Way LW, Stewart L, Gantert W, et al. (2003). Causes and Prevention of Laparoscopic Bile Duct Injuries: Analysis of 252 Cases From a Human Factors and Cognitive Psychology Perspective. Annals of Surgery 237(4): 460–469. https://doi.org/10.1097/01.SLA.0000060680.92690.E9

Weiser TG, Haynes AB, Molina G, et al. (2015). Estimate of the global volume of surgery in 2012: an assessment supporting improved health outcomes. Lancet 385(Suppl 2): S11. https://doi.org/10.1016/S0140-6736(15)60806-6

Yengera G, Mutter D, Marescaux J, Padoy N. (2018). Less is more: Surgical phase recognition with less annotations through self-supervised pre-training of CNN-LSTM networks. https://arxiv.org/abs/1805.08569

Yule S, Flin R, Maran N, Rowley D, Youngson G, Paterson-Brown S. (2008). Surgeons' non-technical skills in the operating room: reliability testing of the NOTSS behavior rating system. World Journal of Surgery 32(4): 548–556. https://doi.org/10.1007/s00268-007-9320-z

Zenati MA, Kennedy-Metz L, Dias RD. (2020). Cognitive Engineering to Improve Patient Safety and Outcomes in Cardiothoracic Surgery. Seminars in Thoracic and Cardiovascular Surgery 32(1): 1–7. https://doi.org/10.1053/j.semtcvs.2019.10.011

van Leeuwen KG, Hedderich DM, Harvey H, Schalekamp S. (2024). How AI should be used in radiology: assessing ambiguity and completeness of intended use statements of commercial AI products. Insights into Imaging 15: 51. https://doi.org/10.1186/s13244-024-01616-9